\documentclass[aps,showpacs,preprint,nofootinbib,preprintnumbers,amsmath,amssymb,graphics]{revtex4}
\usepackage{graphicx}
\input epsf.sty

\begin{document}

\title{Phantom and inflation scenarios from a 5D vacuum through form-invariance transformations of the Einstein equations}
\author{$^{1}$ Maria Laura Pucheu and $^{1,2}$ Mauricio
Bellini\footnote{E-mail address: mbellini@mdp.edu.ar} }
\affiliation{$^{1}$ Departamento de F\'isica, Facultad de Ciencias
Exactas y Naturales, Universidad Nacional de Mar del Plata, Funes
3350, C.P. 7600, Mar del Plata, Argentina. \\ \\ $^{2}$ Instituto
de F\'{\i}sica de Mar del Plata (IFIMAR), \\
Consejo Nacional de Investigaciones Cient\'ificas y T\'ecnicas
(CONICET),\\
Mar del Plata, Argentina.}

\begin{abstract}
We study phantom and inflationary cosmologies using
form-invariance transformations of the Einstein equations with
respect to $\rho$, $H$, $a$ and $p$, from a 5D vacuum. Equations
of state and squared fluctuations of the inflaton and phantom
fields are examined.
\end{abstract}

\pacs{04.20.Jb, 11.10.kk, 98.80.Cq} \maketitle

\section{Introduction}

Currently the universe undergoing to a period of accelerating
expansion. The implications for cosmology should be that the
cosmological fluid is dominated by some sort of fantastic energy
density, which has negative pressure and plays an important role
today. Experimental evidence suggests that the present values of
the dark energy and matter components, in terms of the critical
density, are approximately $\Omega_{\chi}\simeq 0.7$ and
$\Omega_{M} \simeq 0.3$\cite{turner}. The most conservative
assumption is that $\Omega_\chi$ corresponds to a cosmological
parameter which is constant and the equation of state is given by
a constant $\omega _{\chi}= { p}/\rho = -1$, describing a vacuum
dominated universe with pressure $p$ and energy density $\rho$.
Such exotic fluids may be framed in theories with matter fields
that violate the weak energy condition\cite{u}. These models were
called phantom cosmologies, and their study represents a currently
active area of research in theoretical cosmology\cite{uu}. The
motivation for phantom matter is provided by string
theory\cite{..}.

The possibility that our world may be embedded in a
$(4+d)$-dimensional universe with more than four large dimensions
has attracted the attention of a great number of physics. One of
these higher-dimensional theories, where the cylinder condition of
the Kaluza-Klein theory\cite{kk} is replaced by the conjecture
that the ordinary matter and fields are confined to a 4D subspace,
usually referred to as a brane is the Randall and Sundrum
model\cite{rs}. One important physical problem in
higher-dimensional theories is to develop a full understanding of
implications in 4D. Therefore, it is essential to compare and
contrast the effective pictures generated in 4D by different
versions of 5D relativity, where the extra dimension is not
assumed to be compactified. Extensions of General Relativity to
five and more dimensions seem to provide a possible route to
unification of gravity with interactions of particle
physics\cite{a}. The Campbell-Magaard\cite{cm} theorem serves as a
ladder to go between manifolds whose dimensionality differs by
one. This theorem, implies that every solution of the 4D Einstein
equations with arbitrary energy-momentum tensor can be embedded,
at least locally, in a solution of the 5D Einstein field equations
in vacuum.

In this work we explore phantom and inflationary cosmologies using
form-invariance transformations in the Einstein equations. These
transformations were previously studied in this context
in\cite{FI}, but on an extended 5D de Sitter expansion. In this
work we focus our study on an universe which is governed by a
decaying cosmological function. The dynamics of such a universe
was studied previously in \cite{GC}, but in another framework. Our
main interest in this work relies in the study of the dynamics of
the quantum fluctuations of the phantom and the inflaton fields.

\section{Form-invariance transformations: inflation and phantom cosmologies}

Several cosmological models have been pondered in terms of the form-invariance of their 
dynamical equations under a group of symmetry transformations that
preserve the form of the Einstein equations. A form-invariance
transformation is a prescription to relate the quantities $a$,
$H$, $\rho$ and $p$ of our original model, with the quantities
$\bar a$, $\bar H$, $\bar\rho$ and $\bar p$ of what we will call
the transformed model, so that it satisfies
\begin{eqnarray}
&& 3 \bar H^2 = \bar\rho, \\
&& \dot{\bar\rho} + 3 \bar H(\bar\rho+\bar p)=0,
\end{eqnarray}
and are given by
\begin{eqnarray}
&& \bar\rho = \bar\rho(\rho), \\
&& \bar H = \left(\frac{\bar\rho}{\rho}\right)^{1/2} H, \\
&& \bar p = - \bar\rho + \left(\frac{\rho}{\bar\rho}\right)^{1/2}
(\rho + p) \frac{d\bar\rho}{d\rho},
\end{eqnarray}
where $\bar\rho = \bar\rho(\rho)$ is an invertible arbitrary
function. For a barotropic equation of state of a perfect fluid $p
= (\gamma -1) \rho$, one obtains
\begin{equation}
\bar\gamma = \left(\frac{\rho}{\bar\rho}\right)^{3/2}
\frac{d\bar\rho}{d\rho} \gamma.
\end{equation}
We shall consider invariance under these transformations, so that inflationary cosmological scenarios from 
non-inflationary ones can be derived \cite{Chimento1,Chimento2}. Another feature of form-invariance 
transformations is that they can be used also to construct phantom scenarios from cosmologies where the expansion
of the 
universe is governed by a real scalar field $\phi$ \cite{Chimento3}. A form-invariance transformation relates the
scale factor 
$a$, the Hubble parameter $H$, the energy density $\rho$ and the pressure $p$ of the initial cosmological model,
with the 
cosmological parameters $\bar{a}$, $\bar{H}$, $\bar{\rho}$ and $\bar{p}$ of the so called transformed model. Such
a transformation 
leaves invariant the form of the Friedmann dynamical equations of both models. As it was shown in \cite{Chimento3},
under the 
transformation $\bar{\rho}=n^{2}\rho$ with $n$ being a constant,
the scalar field and its potential transform as
\begin{eqnarray}
\dot{\bar\phi}^{2}&=&n\dot{\phi}^{2},\nonumber \\
\bar{V}&=&n^{2}\left(\frac{1}{2}\dot{\phi}^{2}+V\right)-\frac{n}{2}\dot{\phi}^{2},\nonumber
\end{eqnarray}
where $\bar{V}$ is the phantom potential. Thus it can be easily proved that when $n=\pm 1$, the $(+)$ branch
corresponds 
to an identical transformation whereas the $(-)$ branch leads to a
phantom cosmology. In that case the previous expressions become
\begin{eqnarray}
\dot{\bar\phi}^{2}&=&-\dot{\phi}^{2},\nonumber\\
\bar{V}(\bar\phi)&=&\dot{\phi}^{2}+V(\phi). \nonumber
\end{eqnarray}
Clearly the energy density of the transformed model $\bar\rho 
=(1/2)\dot{\bar\phi}^{2}+\bar{V}(\bar\phi)=-(1/2)\dot{\phi}^{2}+\bar{V}(\bar\phi)$ corresponds to a phantom
scalar field 
$\bar\phi$ with the relation $\bar\phi =i\phi$ being valid. In
this letter we shall consider inflation as the initial
cosmological model and phantom cosmology as the transformed one.
Hence, by $p$, $\rho$, $a$ and $H$ ($\bar p$, $\bar\rho$, $\bar a$
and $\bar H$) we shall denote respectively pressure, energy
density, the scale factor and the Hubble parameter, during
inflation (phantom cosmology).\\

In the following section we shall explore the dynamics of the
inflaton and phantom fields through the form-invariance
transformations, but from the 5D vacuum state.\\

\section{Dynamics of the inflaton and phantom fields}

We consider the 5D action
\begin{equation}\label{accion}
I=\int d^{5}x\sqrt{\left|\frac{^{(5)} g}{^{(5)}
g_0}\right|}\left(\frac{R}{16\pi G}+{\cal
L}^{(5)}_{\varphi}\right),
\end{equation}
for a massless scalar field $\varphi$, which is free of any
interactions
\begin{equation}
{{\cal
L}^{(5)}}(\varphi,\varphi_{,B})_{\varphi}=\frac{1}{2}g^{AB}\varphi_{,A}\varphi_{,B}.
\end{equation}
Here, $^{(5)} g$ is the determinant of the covariant tensor metric
$g_{AB}$ \footnote[1]{Latin indices $A,B$ run from $0$ to $4$.},
such that the Riemann-flat metric $R_{ABCD=0}$ is described by
line element\cite{eu}
\begin{equation}\label{metrica}
ds^2=\psi^2\frac{\Lambda(t)}{3}dt^2-\psi^2e^{2\int\\\sqrt{\Lambda(t)/3}dt}dr^2-d\psi^2,
\end{equation}
where $dr^2=dx^2+dy^2+dz^2$, $t$ is the cosmic time, $\Lambda(t)$
is a decreasing cosmological function and $\psi$ a noncompact
extra dimension.

\subsection{The 5D inflaton field dynamics}

The equation of motion for the inflaton field is
\begin{equation}\label{ec.inflaton}
\ddot{\varphi}+\left(3\sqrt{\frac{\Lambda}{3}}-\frac{\dot{\Lambda}}{2\Lambda}\right)\dot{\varphi}-\frac{\Lambda}{3}
e^{-2\int\\\sqrt{\Lambda(t)/3}dt}\nabla_{r}^2\varphi-\frac{\Lambda}{3}\left(4\psi\frac{d\varphi}{d\psi}
+\psi^2\frac{d^2\varphi}{d\psi^2}\right)=0,
\end{equation},
which describes the dynamics on the 5D metric (\ref{metrica}).
\\

\subsection{The 5D phantom field dynamics}

Now we can make the Wick transformation $\bar{\varphi}=i\varphi$,
so that action (\ref{accion}) can be written as
\begin{equation}\label{accion1}
I=\int d^{5}x\sqrt{\left|\frac{^{(5)} g}{^{(5)}
g_0}\right|}\left(\frac{R}{16\pi G}+{\cal
L}^{(5)}_{\bar\varphi}\right),
\end{equation}
where the density Lagrangian written in terms of the massless
scalar field $\bar\varphi$, is
\begin{equation}
{{\cal
L}^{(5)}}(\bar\varphi,\bar\varphi_{,B})_{\bar\varphi}=-\frac{1}{2}g^{AB}\bar\varphi_{,A}\bar\varphi_{,B}.
\end{equation}
The dynamics for the phantom field is given by
\begin{equation}\label{ec.phantom}
\ddot{\bar\varphi}-\left(3\sqrt{\frac{\Lambda}{3}}-\frac{\dot{\Lambda}}{2\Lambda}\right)\dot{\bar\varphi}
+\frac{\Lambda}{3}
e^{-2\int\\\sqrt{\Lambda(t)/3}dt}\nabla_{r}^2{\bar\varphi}+\frac{\Lambda}{3}\left(4\psi\frac{d{\bar\varphi}}{d\psi}
+\psi^2\frac{d^2{\bar\varphi}}{d\psi^2}\right)=0.
\end{equation}
This equation describes the dynamics of the phantom field
$\bar\varphi$ on the metric (\ref{metrica}).\\

\section{Effective 4D dynamics: static foliation}

We consider the equation of motion for the inflaton field
(\ref{ec.inflaton}). This equation can be evaluated on the
foliation $\psi=\psi_0$. The effective 4D dynamics
\begin{equation}\label{4d}
\left.\ddot{\varphi}+\left(3\sqrt{\frac{\Lambda}{3}}-\frac{\dot{\Lambda}}{2\Lambda}\right)\dot{\varphi}-\frac{\Lambda}{3}
e^{-2\int\\\sqrt{\Lambda(t)/3}dt}\nabla_{r}^2\varphi-\frac{\Lambda}{3}\left(4\psi\frac{d\varphi}{d\psi}
+\psi^2\frac{d^2\varphi}{d\psi^2}\right)\right|_{\psi=\psi_0}=0,
\end{equation}
where, after making separation of variables, we can perform the
identification\cite{eu}:
\begin{equation}\label{potencial}
-\left.\frac{\Lambda}{3}\left(4\psi\frac{d\varphi}{d\psi}
+\psi^2\frac{d^2\varphi}{d\psi^2}\right)\right|_{\psi=\psi_0}=\frac{\Lambda
m^2}{3}\,\varphi(t,\vec{r},\psi_0),
\end{equation}
$m^2\geq 0$ being a constant of separation. Hence, the dynamics on
the effective 4D hypersurface
$ds^2=\psi^2_0\frac{\Lambda(t)}{3}dt^2-\psi^2_0
e^{2\int\\\sqrt{\Lambda(t)/3}dt}dr^2$, will be

\begin{equation}\label{4d.inflaton}
\ddot{\varphi}+\left(3\sqrt{\frac{\Lambda}{3}}-\frac{\dot{\Lambda}}{2\Lambda}\right)\dot{\varphi}-\frac{\Lambda}{3}
e^{-2\int\\\sqrt{\Lambda(t)/3}dt}\nabla_{r}^2\varphi +
\frac{\Lambda m^2}{3}\,\varphi=0.
\end{equation}
In the same manner we can obtain the effective equation of motion
for $\bar\varphi$ on the foliation $\psi=\psi_0$, by using
separation of variables on the metric (\ref{ec.phantom}), so that
we obtain
\begin{equation}\label{4d.phantom}
\ddot{\bar\varphi}-\left(3\sqrt{\frac{\Lambda}{3}}-\frac{\dot{\Lambda}}{2\Lambda}\right)\dot{\bar\varphi}
+\frac{\Lambda}{3}
e^{-2\int\\\sqrt{\Lambda(t)/3}dt}\nabla_{r}^2{\bar\varphi}-\frac{\Lambda
m^2}{3}\,{\bar\varphi}=0.
\end{equation}
In order to describe the dynamics of the fields $\varphi$ and
$\bar\varphi$ we shall consider a semiclassical expansion for both
fields: $\varphi =
\varphi_c(t,\psi_0)+\delta\varphi(t,\vec{r},\psi_0)$ and
$\bar\varphi =
\bar\varphi_c(t,\psi_0)+\bar{\delta\varphi}(t,\vec{r},\psi_0)$.

\subsection{Classical dynamics and equations of state}

\begin{figure*}
\includegraphics{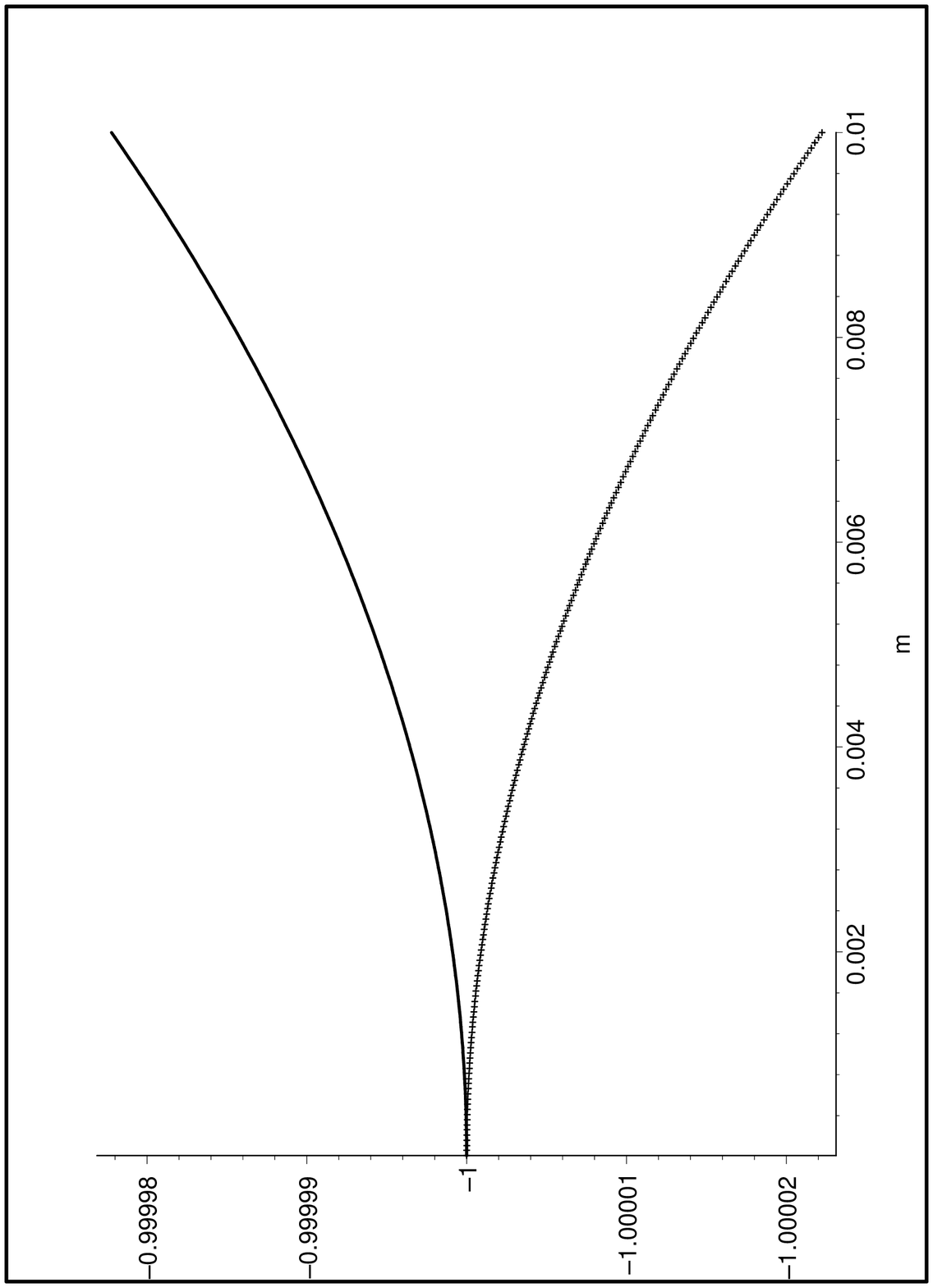}\caption{\label{figura} Values of
${\bar\omega}$ (pointed line) and $\omega$ (continuous line)
plotted for small values of $m$.}
\end{figure*}

We shall consider the case where
\begin{equation}\label{lambda.caso}
\Lambda(t)=\frac{3n^2}{t^2},
\end{equation}
so that $\dot{\Lambda}(t)=-\frac{6n^2}{t^3}$, where $n \gg 1$ is a
constant. The effective scalar potential $V(\varphi)$ is given by
\begin{equation}
V_{eff}(\varphi) = \frac{\Lambda \psi^2}{3}\left.\frac{1}{2}
\left(\frac{d\varphi}{d\psi}\right)^2\right|_{\psi=\psi_0} =
\frac{M^2_{eff}(\psi_0)}{2}\, \varphi^2(t,\vec r,\psi_0),
\end{equation}
where the effective mass is
\begin{equation}
M^2_{eff}(\psi_0) = \frac{\Lambda m^2 }{3}.
\end{equation}
The equation of motion for $\varphi_c(t,\psi_0)$ is
\begin{equation}\label{4di}
\ddot{\varphi_c}+\left(3\sqrt{\frac{\Lambda}{3}}-\frac{\dot{\Lambda}}{2\Lambda}\right)\dot{\varphi_c}
+ \frac{\Lambda m^2}{3}\,\varphi_c=0.
\end{equation}
A particular solution for this equation is
\begin{equation}
\varphi_c(t,\psi_0) = \varphi_0\, t^c,
\end{equation}
where $c$ is given by
\begin{equation}
c = n \left[ -\frac{3}{2} + \sqrt{\frac{9}{4} - m^2}\right].
\end{equation}
The kinetic component of energy density $T_{\varphi_c}$, and the
potential $V_{\varphi_c}$, are
\begin{equation}
T_{\varphi_c}= \frac{1}{2} \left( \frac{3}{\psi_0^2 \Lambda}
\dot\varphi^2_c\right), \qquad V_{\varphi_c}=
\frac{M^2(\psi_0)}{2} \varphi^2_c,
\end{equation}
so that the pressure, $p_{\varphi_c}=T_{\varphi_c}-V_{\varphi_c}$,
and the energy density,
$\rho_{\varphi_c}=T_{\varphi_c}+V_{\varphi_c}$, give us the
equation of state for inflation: $\omega=p_{\varphi_c}
/\rho_{\varphi_c}$, which in this case is given by
\begin{equation}\label{omega sin aprox2}
\omega=\frac{\left(-\frac{3}{2}+\sqrt{\frac{9}{4}-m^2}\right)^2-m^2}
{\left(-\frac{3}{2}+\sqrt{\frac{9}{4}-m^2}\right)^2+m^2}.
\end{equation}
On the other hand, the effective mass for the phantom field
is
\begin{equation}\label{masa.fantasma.estatica}
\bar{M}^2_{eff}(\psi_0)=-\frac{\Lambda
\psi^2_0}{3}\left(\frac{2c^2}{t^2}+\frac{2m^2}{\psi_0^2}\right).
\end{equation}
Notice that $\lim_{t\rightarrow
\infty}{\bar{M}^2_{eff}(\psi_0)}\rightarrow -2 M^2_{eff}$. The
equation of state for the phantom system
$\bar{\omega}_{\varphi_c}=\bar{p}_{\varphi_c}
/\bar{\rho}_{\varphi_c}$, is
\begin{equation}\label{omega fantasma sin aprox.}
\bar{\omega}=-\frac{3\left(-\frac{3}{2}+\sqrt{\frac{9}{4}-m^2}\right)^2+m^2}
{\left(-\frac{3}{2}+\sqrt{\frac{9}{4}-m^2}\right)^2+m^2},
\end{equation}
where we have taken into account that $\bar{\varphi}_c =
i\,\varphi_c$, and that
\begin{equation}
\bar{p}_{\varphi_c} = \frac{\dot{\varphi^2_c}}{2} -
V(\varphi_c),\qquad \bar{\rho}_{\varphi_c} =
-\frac{\dot{3\varphi^2_c}}{2} - V(\varphi_c).
\end{equation}
To slow-roll conditions to be fulfilled, we would need that $m^2
\ll 1$, so that $\omega \gtrsim -1$ and $\bar\omega < -1$, for
inflation and phantom cosmology, respectively. It is interesting
to notice that in both cases $\lim_{m\rightarrow
0}{\omega}=\lim_{m\rightarrow 0}{\bar\omega}=-1$. Finally, values
of ${\bar\omega}$ (pointed line) and $\omega$ (continuous line)
were plotted for small values of $m$ in the figure (\ref{figura}).
Notice that both are very symmetric with respect to $-1$, and they
are very sensitive to $m$.\\

\subsection{Quantum fluctuations}

Both equations (\ref{4d.inflaton}) and (\ref{4d.phantom}) are
linear, so that can be expanded in terms of Fourier modes. The
dynamics of the time-dependent modes of quantum fluctuations in
both cases, are respectively given by
\begin{equation}\label{equ1}
\ddot{\delta\varphi}_k+\left(3\sqrt{\frac{\Lambda}{3}}-\frac{\dot{\Lambda}}{2\Lambda}\right)\dot{\delta\varphi}_k
+\left[\frac{k^2
\Lambda}{3} e^{-2\int\\\sqrt{\Lambda(t)/3}dt} + \frac{\Lambda
m^2}{3}\right]\,\delta\varphi_k=0,
\end{equation}
and
\begin{equation}\label{equ2}
\ddot{\bar{\delta\varphi}}_k-\left(3\sqrt{\frac{\Lambda}{3}}-\frac{\dot{\Lambda}}{2\Lambda}\right)
\dot{\bar{\delta\varphi}}_k+\left[\frac{k^2 \Lambda}{3}
e^{-2\int\\\sqrt{\Lambda(t)/3}dt}- \frac{\Lambda
m^2}{3}\right]\,{\bar{\delta\varphi}}_k=0.
\end{equation}
If we take into account the case (\ref{lambda.caso}), the general
solutions of the equations (\ref{equ1}) and (\ref{equ2}), are
respectively
\begin{eqnarray}
&& \delta\varphi_k(t) = \left(\frac{t}{t_0}\right)^{-3n/2}\,\left[C_1\,{\cal H}^{(1)}_\nu[x(t)]
+C_2\,{\cal H}^{(2)}_\nu[x(t)]\right] , \\
&& \bar{\delta\varphi}_k(t) =
\left(\frac{t}{t_0}\right)^{3n/2+1}\, \left[\bar{C}_1\,{\cal
H}^{(1)}_{\mu}[x(t)]+\bar{C}_2\, {\cal H}^{(2)}_\mu[x(t)]\right] ,
\end{eqnarray}
such that ${\cal H}^{(1,2)}_{\nu}$ are the first and second kind
Hankel functions, $\nu ={3\over 2}\sqrt{1-4m^2/9}$, $\mu ={3\over
2}\sqrt{1+{4m^2\over 9}+{12\over 9n}+{4\over 9n^2}}$ and $x(t)=k
\left(t/t_0\right)^{-n}$. Notice that, since $m^2 \ll 1$ and $n
\gg 1$, we obtain that $\nu < 3/2$ and $\mu > 3/2$.

In order to calculate the squared $\delta\varphi$-fluctuations on
cosmological scales $k \ll k_0(t)$, we shall need the asymptotic
expressions of the Hankel functions for $x \ll 1$: $\left.{\cal
H}_\nu^{(2)}[x(t)]\right|_{x\ll 1}\simeq -{i\over
\pi}\Gamma(\nu)\left({x\over 2}\right)^{-\nu}$. These fluctuations
describe the infrared sector, which during inflation is given by
super Hubble fluctuations, and are given by
\begin{equation}
\left.\left<\delta\varphi^2\right>\right|_{IR}=\frac{H^3}{\pi^3}\frac{B
B^{*}\,2^{2\nu-2}}{(3-2\nu)} \left[\epsilon
\left(\frac{n^2}{3}\left(\frac{9}{4}-\frac{1}{4n^2}+m^2\right)\right)^{1/2}\right]^{3-2\nu},
\end{equation}
where $B={i\over 2} \sqrt{{\pi\over H}}=C_{2}/H^{3/2}$ and
$H=n/t_0$, $t_0$ being the time at the end of inflation. The modes
are normalized by choosing the Bunch-Davies vacuum. Notice that
for $k \gg k_0(t)$
\begin{equation}
{k_0}^2=\frac{n^2}{3}\left(\frac{9}{4}-\frac{1}{4n^2}+m^2\right)\,t^{2n},
\end{equation}
the solutions become unstables. This function is the time
dependent wavenumber related to the size of the horizon.

The $\bar\varphi$-fluctuations in phantom cosmology hold
\begin{equation}
\left.\left<\bar{\delta\varphi}^2\right>\right|_{IR}=\frac{H^3}{\pi^3}\frac{\bar{B}
\bar{B}^{*}\,2^{2\mu-2}}{(3-2\mu)} \left[\epsilon
\left(\left(\frac{(3n+1)(3n+3)}{4n^2}+m^2\right)\right)^{1/2}\right]^{3-2\nu}
\left(\frac{t}{t_0}\right)^{6n+2},
\end{equation}
which increases with time. Here, $\bar{B}={i\over 2}
\sqrt{{\pi\over H}}=\bar{C}_{2}/H^{3/2}$ which is determined by
normalization of the $\delta\bar{\varphi}_k(t)$ modes.

The relevant wavenumber $\bar{k}_0$ is
\begin{equation}
\bar{k}_0^2=\left(\frac{(3n+1)(3n+3)}{4n^2}+m^2\right)\,t^{2n}.
\end{equation}
This case is similar to whose of the inflaton field, because in
both cases the modes are unstables on cosmological scales and
oscillate on sub Hubble scales. Finally, the squared
$\dot{\bar{\delta\varphi}}$-fluctuations increase as
$\left.\left<\dot{\bar{\delta\varphi}}^2\right>\right|_{IR}\sim
t^{6n}$, so that they can be neglected with respect to
$\left.\left<\bar{\delta\varphi}^2\right>\right|_{IR}$, which
remains dominant at the end of
the phantom epoch.\\

\section{Final Comments}

We have revisited phantom and inflationary scenarios through the
form-invariance transformations of the Einstein equations with
respect to $\rho$, $H$, $a$ and $p$, from a 5D vacuum state. In
particular, the dynamics of quantum fluctuations of the phantom
and the inflaton fields was studied. One of the interesting
results here obtained is that the behavior of phantom field
fluctuations $\left<\bar\varphi^2\right>$ is different to the
inflaton ones; the large-scale (super Hubble) amplitude of these
fluctuations is not freezed and increases dramatically with time.
Hence, phantom fields should be not a good candidate to describe
inflation, at least under the conditions here studied. However, we
believe that this field is a good candidate to describe the
evolution of a asymptotic future of the universe. Finally, we have
studied the equation of state for inflation and phantom scenarios.
The results plotted in figure (\ref{figura}) show that the big rip
observed in phantom cosmology
is sensitive to the parameter $m$.\\

\section*{Acknowledgements}

\noindent M.B. acknowledges UNMdP and CONICET Argentina for
financial support. M.L.P. acknowledges UNMdP for financial
support.\\

\end{document}